\magnification=1200\hsize=15.6 truecm\vsize=22.4 truecm

\def\be{ $$ } \def\fe{ $$ } \def\eqn{ \eqno }
\def\cite{  } \def\ref{  }

\def\spose#1{\hbox to 0pt{#1\hss}}
\def\Libra{\spose {\raise 0.6pt \hbox{{--}}} {\cal L}}
\def\Diam{\spose {\raise 0.3pt\hbox{+}} {\diamondsuit} }

\def\sqr#1#2{{\vcenter{\hrule height.4pt\hbox{\vrule width.8pt height#2pt
\kern#1pt\vrule width.8pt}\hrule height.4pt}}}
\def\Square{\mathchoice{\sqr78\,}{\sqr78\,}\sqr{20.0}{18}\sqr{20.0}{18}}

\font\srm=cmr6
\font\Srm=cmr8

\centerline{\it 1982 Les Houches Lecture Notes}
\vskip 1 cm
\centerline{INTERACTION OF GRAVITATIONAL WAVES}
\centerline{WITH AN ELASTIC SOLID MEDIUM}
\bigskip
\centerline{ B. Carter}
\bigskip
\centerline{ Groupe d'Astrophysique Relativiste}
\centerline{ Observatoire de Paris}
\centerline{ 92 Meudon, France}
\vskip 1.2 cm

\centerline{$\rm\underline{Contents}$}
\smallskip\parindent = 1.6 cm

1. Introduction.

2. Kinematics of a Material Medium: Material Representation.

3. Kinematics of a Material Medium: Convected Differentials.

4. Kinematics of a Perfect Elastic Medium.

5. Small Gravitational Perturbations of an Elastic Medium.

\vskip 1 cm
\parindent =0.6 cm

\bigskip
\noindent
1 INTRODUCTION
\medskip

Although the presently most promising approach to the direct detection
of gravitational radiation would seem to be provided by the use of
appropriate electromagnetic field configurations (e.g. that of and
optical interferometer) the original, and until now most highly
developed method has been based on the use of an elastic solid
configuration, traditionally a cylindrical bar of the kind originally
introduced by Weber. A highly simplified but essentially adequate
description of the interaction of gravitational radiation with such
an apparatus was given by Weber himself in his seminal 1960 paper
\cite{[1]}. For a more detailed analysis, and for the treatment of more
general configurations such as that of the earth as a whole, a system
of wave equations governing the interaction of weak gravitational
radiation with an elastic solid was derived soon after by Dyson
\ref{[2]}, Papapetrou \ref{[3]}, and others, subject to the limitation
that non-linearities due to self gravitation of the solid medium are
neglected, as is entirely justifiable in a terrestrial context.

While the final outcome of the present article will be a rederivation
of the weak field wave equations to which we have just referred, we shall
nevertheless use a quite different approach to that of Dyson and
Papapetrou. Instead of working throughout in an only approximately
self consistent linearised scheme, as they did, we shall first
set up the exactly self consistent fully non-linear theory of the
interaction of a gravitational field with an elastic solid in accordance
with Einsein's theory. The general, non- linear theory is in any case
needed for application to the more exotic context of neutron star
deformations, as was discussed e.g. by Carter and Quintana \ref{[4]}.
Starting from this mathematically sound basis, we shall then proceed
to  the derivation of the weak field limit in two successive stages
of approximation: we shall first impose the restriction that the
gravitational radiation be weak, even though the unperturbed background
field may still be strong (as in the case of a neutron star); finally
we shall impose the condition that the background field also be
weak (as in the case of the earth) so as to obtain the Dyson Papapetrou
equations. 

Unlike the treatments used by other authors such as those of the schools
of Eringen (see e.g. Maughin \ref{[5]}) or of Souriau (who was the
first to set up the fully non-linear elasticity theory used here
\ref{[6]}) our present treatment will be based on the use of
convected differentials, a powerful technique for the analysis of
material media (elastic or otherwise) that generalises the more
restricted convected differentiation procedure first introduced by
Oldroyd \ref{[7]}. In providing a self-contained introduction to the
general concept of convected differentals and differentiation, and
to their application to the theory of an elastic solid and its
interaction with gravitation, the present article condenses several
previous publications describing work carried out in collaboration
with H. Quintana \ref{[8]}, \ref{[9]}, \ref{[10]}, \ref{[11]},
\ref{[12]}. This article thus serves to update an earlier
introductory survey by Ehlers \ref{[13]} .

\bigskip
\noindent
2 KINEMATICS OF A MATERIAL MEDIUM: MATERIAL REPRESENTATION

\medskip

In Newtonian theory a $\rm\underline{material}$ $\rm\underline{medium}$ is 
usually visualised as a three dimensional manifold whose configuration at a 
given instant is specified by a non-singular mapping into three dimensional 
Euclidean space, the motion of the medium being given by the time variation 
of the maping. However in General Relativity theory, where there is in 
general no canonically preferred set of three-dimensional sections 
(Euclidean or otherwise) of the fundamental four-dimensional spacetime 
manifold ${\cal M}$ say, it is more convenient and natural to proceed the 
other way about. The medium itself is still to be conceived in the abstract 
as a three-dimensional manifold, ${\cal X}$ say, (whose points represent 
idealised particles of the material) but its motion can be specified by a 
necessarily degenerate (four to three dimensional) mapping, ${\cal P}$ say, 
of the world-tube traversed by the matter in ${\cal M}$ onto the manifold 
${\cal X}$, the world line of each idealised particle of the medium being 
projected under ${\cal P}$ onto the corresponding point in ${\cal X}$.

We shall use Greek indices $\mu, \nu, ... $ to specify ordinary
spacetime tensors on ${\cal M}$, and capital Roman indices 
{\Srm A, B,} ...  to specify tensors on the manifold ${\cal X}$
representing the medium. The spacetime indices $\mu, \nu, ... $ may
be thought of in the traditional manner as specifying tensor components
defined in terms of a local system of coordinates $x^\mu$ ($\mu= 0,1,2,3$)
on ${\cal M}$ while similarly the material indices {\Srm  A, B, ...}
may be thought of as defined in terms of local coordinates
$X^{\hbox{\srm A}}$ ({\Srm A}=1,2,3)  on ${\cal X}$. However, following 
Penrose\cite{[14]}, it wil also be comvenient to interpret the index symbols
in the abstract manner as merely an indication of the tensorial
(or more general) character of the quantities concerned (rank, 
co/contravariant quality, and in the present case association with
${\cal M}$ or ${\cal X}$) rather than as integers specifying concrete 
components. Thus with the indices interpreted in this loose sense, the
statement that a point with coordinates $X^{\hbox{\srm A}}$ in ${\cal X}$ is 
the image under ${\cal P}$ of a point with coordinates $x^\mu$ in ${\cal M}$ 
may legitimately be expressed in a simple and natural manner by
\be X^{\hbox{\srm A}}={\cal P}\{x^\mu\} \eqn{(2.1)}\fe
whereas with a strict traditional interpretation of {\Srm A} and $\mu$ such
an equation would be nonsense.

The operation ${\cal P}$ of projection from ${\cal M}$ onto ${\cal X}$
evidently induces a corresponding projection, which we shall also denote 
by ${\cal P}$, from tangent vectors at any given point $x^\mu$ in ${\cal M}$
onto tangent vectors at the corresponding point ${\cal P}\{x^\mu\}$
in ${\cal X}$. In an obviously natural notation we shall denote the projected 
image of a tangent vector by the same basic symbol, distinguishing the
image from the original vector only by the appropriate change from Roman to 
Greek of the index symbol, i.e. for any spacetime tangent vector $\xi^\mu$
we shall set
\be \xi^{\hbox{\srm A}}={\cal P}\{\xi^\mu\} \, .\eqn{(2.2)}\fe
Due to the degeneracy of ${\cal P}$ there is in general no corresponding 
induced projection operation for covectors except in the particular case
of a covector $\alpha_\mu$ that is orthogonal to the congruence of 
worldlines in the sense that
\be u^\mu \alpha_\mu=0 \, ,\eqn{(2.3)}\fe
where  the vector $u^\mu$ is tangent to the worldline at the spacetime
point in question. In this particular case the projection will induce a 
corresponding covector on ${\cal X}$, for which as before we shall use 
the same basic symbol, i.e. we shall write 
\be \alpha_{\hbox{\srm A}}={\cal P}\{\alpha_\mu\}\, ,\eqn{(2.4)}\fe
and in this case the operation is reversible, i.e. we shall have a bijection
\be \alpha_\mu\, \leftrightarrow\, \alpha_{\hbox{\srm A}}\eqn{(2.5)}\fe
between orthogonal covectors at $x^\mu$ in ${\cal M}$ and covectors at
${\cal P}\{x^\mu\}$ in ${\cal X}$. We can use the (pseudo-) metric
tensor $g_{\mu\nu}$ on the spacetime manifold ${\cal M}$ (with signature
$- + + +$ and units such that $c=1$) to define a corresponding bijection 
for tangent vectors by restricting our attention to tangent vectors on 
${\cal M}$ that are orthogonal to the world lines in the metric sense, i.e.
\be u_\mu\xi^\mu=0\, ,\hskip 1 cm u_\mu=g_{\mu\nu}u^\nu \, .\eqno{(2.6)}\fe
Subject to this restriction the projection \ref{(2.1)} will have the same
reversibility property as \ref{(2.4)}, i.e.
\be \xi^\mu\, \leftrightarrow\, \xi^{\hbox{\srm A}} \eqn{(2.7)}\fe
in the same sense as \ref{(2.5)}.

The natural 1 - 1 correspondence that has just been defined between vectors
or covectors orthogonal to the worldlines in spacetime and vectors or
covectors in the material medium can evidently be extended directly to 
general orthogonal tensors as defined in terms of tensor products of 
orthogonal vectors and covectors. This correspondence is of vital
importance for setting up any physical theory of the mechanical behaviour
of the medium, since its spacetime evolution must be described in terms of
geometrical quantities (tensors etc.) defined in spacetime, whereas its
intrinsic properties can only be described in terms of quantities described
directly in terms of the manifold ${\cal X}$ representing the material medium
(a requirement commonly dignified as the ``principle of material 
objectivity'').

Since there is no guarantee that all physically relevant spacetime tensors
will turn out to be automatically orthogonal to the worldlines, it is
important to remark that they can always be canonically decomposed into 
a set of orthogonal tensors of equal or lower order by using the orthogonal
projection tensor $\gamma^\mu_\nu$. In terms of the normalised unit tangent
vector $u^\mu$ and the corresponding covector $u_\mu$ defined by
\be u^\mu u_\mu=-1 \eqn{(2.8)}\fe
this orthogonal projection operator is defined in terms of the unit tensor
$g^\mu_\nu$ by
\be\gamma^\mu_\nu=g^\mu_\nu+u^\mu u_\nu\, .\eqn{(2.9)}\fe
In the particular case of an ordinary tangent vector $v^\mu$ the natural
decomposition into a set consisting of an orthogonal part $_\perp\!v^\mu$
and a scalar $v^\Vert$ is given by
\be v^\mu= _\perp\! v^\mu+v^\Vert u^\mu \, ,\eqn{(2.10)}\fe
where
\be _\perp\! v^\mu=\gamma^\mu_\nu v^\nu\, ,\hskip 1 cm
 v^\Vert=-u_\mu v^\mu \, ,\eqn{(2.11)}\fe
while for  a general covector $\omega_\mu$ we similarly have
\be \omega_\mu=_\perp\!\omega_\mu+\omega_\Vert u_\mu \, ,\eqn{(2.12)}\fe
where 
\be_\perp\!\omega_\mu=\gamma^\nu_\mu\omega_\nu\, ,\hskip 1 cm
\omega_\Vert= -u^\mu\omega_\mu\, .\eqn{(2.13)}\fe
This allows us to represent them in terms of sets of corresponding
geometrical quantities defined at the corresponding point in the
material medium in the form
\be v^\mu\, \leftrightarrow\, \{ _\perp\!v^{\hbox{\srm A}}, v^\Vert\}
\eqn{(2.14)}\, ,\fe
and 
\be\omega_\mu\, \leftrightarrow\, \{_\perp\!\omega_{\hbox{\srm A}},
\omega_\mu\}\, ,\eqn{(2.15)}\fe
the extension to general tensors of higher order being straightforeward
albeit cumbersome.

\bigskip
\noindent
3 KINEMATICS OF A MATERIAL MEDIUM: CONVECTED DIFFERENTIALS
\medskip

The concept of a material representation of spacetime tensors in terms 
of sets of tensors defined on the medium space (as defined by \ref{(2.14)}
and \ref{(2.15}) ) gives rise naturally to the concept of $\rm\underline
{material}$ $\rm\underline{variation}$ defined as the difference between 
the material representations in any Lagrangian (i.e. worldline preserving) 
variation between different configurations, which may either arise from a 
mapping between different physically conceivable evolutions of the medium or 
else merely from a time displacement in a single evolution. The spacetime 
tensor corresponding to the infinitesimal material variation between 
nearby states of evolution of the medium will be referred to as the 
convected differential\cite{[12]}. We shall use the symbol $\Delta$ to 
denote an ordinary Lagrangian differential and we shall use the notation 
$d[\,\,  ]$ for a corresponding convected differential. Thus in the case of 
a vector and a covector respectively, \ref{(2.14}) and \ref{(2.15)} give 
rise to the correspondences 
\be d[v^\mu]\, \leftrightarrow \,\{\Delta _\perp\!v^{\hbox{\srm A}},
\Delta v^\Vert\}\, ,\eqn{(3.1)}\fe
and
\be d[\omega]\, \leftrightarrow\, \{\Delta _\perp\!\omega_{\hbox{\srm A}},
\Delta \omega_\Vert\}\, .\eqn{(3.2)}\fe
To evaluate these expressions we use the facrthat the world line preserving
variation can only change the magnitude but not the direction of the unit
tangent vector $u^\mu$: explicitly
\be \Delta u^\mu={_1\over ^2} u^\mu u^\nu u^\rho \Delta_{\rho\sigma}
\eqn{(3.3)}\fe
where we use the abbreviation
\be\Delta_{\rho\sigma}=\Delta g_{\rho\sigma} \eqn{(3.4)}\fe
for the Lagrangian variation of the metric. In the case of a vector we
obtain
\be \Delta v^\Vert= -u_\nu \Delta v^\nu-v^\nu\Delta u_\nu \eqn{(3.5)}\fe
\be \Delta _\perp\!v^\mu=\gamma^\mu_\nu\Delta v^\nu+ v^\nu\big(
u^\mu \Delta u_\nu+ u_\nu\Delta u^\mu\big)\, .\eqn{(3.6)}\fe
Since the bracketed quantity in the last expresion is automatically parallel
to $u^\mu$ it does not contribute to the material projection, so we obtain
\be \Delta _\perp\!v^{\hbox{\srm A}}= {\cal P}\{\Delta _\perp\! v^\mu\}
={\cal P}\{ \gamma^\mu_\nu v^\nu\} \eqn{(3.7)}\fe
and hence
\be d[v^\mu]=\gamma^\mu_\nu\Delta v^\nu+u^\mu\Delta v^\Vert\fe
\be \hskip 1.7 cm =\Delta v^\mu -u^\mu v^\nu\Delta u_\nu \, .\eqn{(3.8})\fe 
In the case of a covector we obtain
\be \Delta\omega_\Vert=-u^\nu\Delta \omega_\nu -\omega_\nu\Delta u^\nu
\eqno{(3.9)}\fe
\be \Delta _\perp\!\omega_\mu=\gamma^\nu_\mu\Delta\omega_\nu +
\omega_\nu\big(u^\nu\Delta u_\mu+u_\mu\Delta u^\nu\big)\, ,\eqn{(3.10)}\fe
both terms in the last expression being automatically orthogonal to $u^\mu$.
Thus using
\be \Delta _\perp\!\omega_{\hbox{\srm A}}={\cal P}\{\Delta _\perp\!
\omega_\mu\} \eqn{(3.11)}\fe
we immediately obtain
\be d[\omega_\mu]=\Delta _\perp\!\omega_\mu+ u_\mu\Delta \omega_\Vert\fe
\be\hskip 1.7 cm =\Delta \omega_\mu+\omega_\nu u^\nu\Delta u_\mu
\, .\eqn{(3.12)}\fe
The extension to a general tensor is now automatic: it suffices to add
an appropriately analogous term for each extra index. Thus for a general
mixed tensor $T^{\mu ...}_{\,\,\nu ...}$ one obtains the 
relation\cite{[12]} between convected and Lagrangian differentials in
the form
\be d[T^{\mu ...}_{\,\,\nu ...}]=\Delta T^{\mu ...}_{\,\,\nu ...}+
T^{\mu ...}_{\,\,\rho...}\, u^\rho\Delta u_\nu + ...\fe
\be \hskip 3 cm - T^{\rho ...}_{\,\,\nu ...}\, u^\mu\Delta u_\rho - ...
\eqn{(3.13})\fe
where the Lagrangian differential of the covector $u_\mu$ is obtainable by
substituting the formula \ref{(3.3}) in the expression
\be \Delta u_\mu= g_{\mu\nu}\Delta u^\nu+ u^\nu\Delta_{\mu\nu}
\, .\eqn{(3.14})\fe

An important particular case\cite{[8]} covered by the general formula
\ref{(3.13)} is that in which instead of making a comparison between
nearby but different states of material motion (as one needs to do in
perturbation theory) one wishes to study $\rm\underline{time}$ variations 
in a single given state of motion. This corresponds to the case in which
the Lagrangian variation is simply given by $\rm\underline{Lie}$ $\rm
\underline{differentiation}$ with respect to a time displacement vector 
field, $\zeta^\mu$ say, which we shall denote by $\zeta\Libra$, i.e. we set
\be \Delta=\zeta\Libra\, ,\eqn{(3.15)}\fe
where the vector $\zeta^\mu$ is an arbitrarily normalised tangent to
the flow, which can therefore be expressed in the form
\be \zeta^\mu= u^\mu\Delta\tau\, ,\eqn{(3.16)}\fe
where $\Delta\tau$ is an arbitrary scalar field interpretable as 
representing the local value of the corresponding infinitesimal proper 
time displacement along the world lines. Although the various terms in
\ref{(3.13)} will involve derivatives of $\zeta^\mu$, the intrinsic
nature of the material variation will ensure that the convected
differential will only depend on the scalar value of the displacement
$d\tau$ at the point under consideration, so that it will take the form
\be d[T^{\mu ...}_{\,\,\nu ...}]= [T^{\mu ...}_{\,\,\nu ...}]\dot{}\,
d\tau \eqn{(3.17)}\fe
where the tensor $[T^{\mu ...}_{\,\,\nu ...}]\dot{}$ so defined is what we 
refer to as the $\rm\underline{convected}$ $\rm\underline{derivative}$. By 
direct substitution of \ref{(3.15)} and \ref{(3.16)} in \ref{(3.13)} one 
can check that the terms involving gradients of $d\tau$ do indeed cancel 
out, so that one recovers the original formula\cite{[8]} for the convected 
derivative, namely
\be [T^{\mu ...}_{\,\,\nu ...}]\dot{}=T^{\mu ...}_{\,\,\nu ...}+
T^{\mu ...}_{\,\,\rho ...}(\dot u_\nu+\nabla_{\!\nu})u^\rho + ...\fe
\be \hskip 3 cm - T^{\rho ...}_{\,\,\nu ...}(\dot u^\rho+\nabla_{\!\rho}
) u^\mu - ... \eqn{(3.18)}\fe
where $\nabla_{\!\mu}$ is the usual metric covariant differentiation
operator and where we use a simple dot withoutsquare brackets to denote
covariant differentiation with respect to the propert time, i.e.
$\dot{}\,=u^\rho\nabla_\rho$. In the particular case of the unit
tangent vector itself, the dot operation gives the acceleration 
vector, whose covariant form is also expressible as the Lie derivative:
\be \dot u_\mu = u\Libra u_\mu \, .\eqn{(3.19)}\fe

Just as \ref{(3.13)} is a generalisation of the earlier formula
\ref{(3.18)}, so also \ref{(3.18)} is itself a generalisation of a
previuos formula given by Oldroyd\cite{[7]} for the particular case of
tensors entirely orthogonal to the world lines, for which, as pointed
out by Ehlers\cite{[13]}, the convected derivative reduces to the
orthogonal projection of the Lie derivative. An important example is
the divergence tensor $\theta_{\mu\nu}$ of the material flow, as defined 
by the decomposition
\be \nabla_{\!\mu} u_\nu=\theta_{\mu\nu}+\omega_{\mu\nu}-\dot u_\mu
u_\nu\, ,\eqn{(3.20})\fe
where the vorticity tensor $\omega_{\mu\nu}$ is antisymmetric and
$\theta_{\mu\nu}$ is symmetric. It is related to the strain tensor
$\gamma_{\mu\nu}$ (i.e. the covariant version of the orthogonal
projection tensor \ref{(2.19)} ) by
\be \theta_{\mu\nu}={_1\over^2}[\gamma_{\mu\nu}]\,\dot{}
\, .\eqn{(3.21)}\fe

\bigskip
\noindent
4 MECHANICS OF A PERFECT ELASTIC MEDIUM
\medskip

The convected differential and derivative that were described in the
previous section are potentially useful for the kinetic analysis of any 
kind of material medium. One of the simplest applications is to the 
theory of a medium whos behaviour satisfies the following (by now 
standard) criterion of perfect elasticity. A perfect
$\rm\underline{elastic}$ $\rm\underline{medium}$ can be characterised 
succinctly by the condition that its energy - momentum tensor is a 
material function of the metric tensor with respect to the flow field 
specified by its timelike eigenvector. This means that the energy 
momentum tensor takes the form
\be T^{\mu\nu}=\rho u^\mu u^\nu+ P^{\mu\nu}\eqn{(4.1)}\fe
where the pressure tensor satisfies the orthogonality condition
\be P^{\mu\nu}u_\nu =0 \, ,\eqn{(4.2)}\fe
so that the material representation of $T^{\mu\nu}$ is expressible as
\be T^{\mu\nu}\,\leftrightarrow\,  \{P^{\hbox{\srm AB}}, 0, \rho\}
\, ,\eqn{(4.3)}\fe
and this representation must be a function of the corresponding 
representation
\be g_{\mu\nu} \, \leftrightarrow\, \{\gamma_{\hbox{\srm AB}}, 0,
-1\}\, .\eqn{(4.4)}\fe
Thus at each point in the three dimensional manifold ${\cal X}$
representing the medium there are well defined functions determining
the pressure components $P^{\hbox{\srm AB}}$ and also the density
$\rho$ in terms of the components $\gamma_{\hbox{\srm AB}}$. There
is however a restriction that prevents these functions from all
being chosen arbitrarily, namely the local energy - momentum
conservation law
\be \nabla_{\!\mu} T^{\mu\nu}=0 \, ,\eqn{(4.5)}\fe
which has four independent components, whereas the acceleration of
the flow has only three independent degrees of freedom. In order to
avoid having an overdetermined system of equations of motion, one
must require thqt the component of ({4.5}) along the flow (i.e. the
conservation of rest frame energy as distinct from momentum) should
be satisfied as an identity. The remaining independent equations are
given by
\be \gamma^\mu_\nu\nabla_{\!\rho} T^{\rho\nu}=0 \, ,\eqn{(4.6)}\fe
which is equivalent to the equations of motion
\be \rho \dot u^\mu = -\gamma^\mu_\nu\nabla_{\!\rho} P^{\rho\nu}
\, .\eqn{(4.7)}\fe
The equation that must be satisfied identically is
\be u_\nu \nabla_{\!\mu} T^{\mu\nu} = 0 \, ,\eqn{(4.8)}\fe
which is equivalent to
\be \dot\rho= -\big(\rho \gamma^{\mu\nu}+ P^{\mu\nu}\big)
\theta_{\mu\nu}\, .\eqn{(4.9)}\fe
Now it follows from \ref{(3.21)} that this last can be expresssed in
terms of convected derivatives as
\be [\rho]\dot{}\,= -{_1\over^2}\big(\rho \gamma^{\mu\nu}+ P^{\mu\nu}
\big) [\gamma_{\mu\nu}]\dot{}\, ,\eqn{(4.10)}\fe
which means that the convected variations allong the world lines must
satisfy
\be d[\rho]\,= -{_1\over^2}\big(\rho \gamma^{\mu\nu}+ P^{\mu\nu}
\big) d [\gamma_{\mu\nu}]\, ,\eqn{(4.11)}\fe
and hence that the corresponding variation of the material projections
in ${\cal X}$ must satisfy
\be  d\rho\,= -{_1\over^2}\big(\rho \gamma^{\hbox {\srm AB}}+ 
p^{\hbox{\srm AB}}\big) \, d\gamma_{\hbox{\srm AB}}\, .\eqn{(4.12)}\fe
This will be satisfied automatically for an arbitrary equation of
state $\rho=\rho\{\gamma_{\hbox{\srm AB}}\}$ if and only if the
corresponding equations for the six algebraically independant
pressure components are specified by
\be P^{\hbox{\srm AB}}= -2{\partial\rho\over\partial
\gamma_{\hbox{\srm AB}} } -\rho\gamma^{\hbox{\srm AB}}\, .\eqn{(4.13)}\fe

By carrying out a second partial differentiation of the single
equation of state function for $\rho$ with respect to the strain
$\gamma_{\hbox{\srm AB}}$ we deduce that the material variation of
the pressure tensor will be given by
\be dP^{\hbox{\srm AB}}= -{_1\over^2}\big( E^{\hbox{\srm ABCD}}
+P^{\hbox{\srm AB}} \gamma^{\hbox{\srm CD}}\big) d\gamma_{\hbox{\srm CD}}
\, ,\eqn{(4.14)}\fe
where the $\rm\underline{elasticity\ tensor}$, whose material projection is
defined by
\be E^{\hbox{\srm ABCD}} = -2{\partial p^{\hbox{\srm AB}}\over
\partial \gamma_{\hbox{\srm CD}}} - p^{\hbox{\srm AB}}
\gamma^{\hbox{\srm CD}} \eqn{(4.15)} \fe
will obey the symmetry conditions
\be E^{\mu\nu\rho\sigma}= E^{(\mu\nu)(\rho\sigma)}= E^{\rho\sigma\mu\nu}
\eqn{(4.16)}\fe
as well as the orthogonality requirement
\be E^{\mu\nu\rho\sigma}u_\sigma =0\, .\eqn{(4.17)}\fe

A familiar special case of a perfectly elastic medium is that of an
ordinary perfect fluid, which can be defined in the present context by
the condition that its density be a function only of the determinant
$|\gamma|$ of the material projection of the strain tensor, i.e.
\be \rho=\rho\{|\gamma|\}\, .\eqn{(4.18)}\fe
It follows from \ref{(4.13)} that the pressure tensor will then take the
purely isotropic form
\be P^{\mu\nu}= P\gamma^{\mu\nu}\, ,\eqn{(4.19)}\fe
with the pressure scalar given by
\be P=-2|\gamma|{d\rho\over d|\gamma|}\, ,\eqn{(4.20)}\fe
while the elasticity tensor will be given in terms of the bulk
modulus
\be \beta= -2|\gamma|{d P\over d |\gamma|}\, ,\eqn{(4.21)}\fe
by the formula
\be E^{\mu\nu\rho\sigma}=(\beta - P)\gamma^{\mu\nu}\gamma^{\rho\sigma}
+ 2 P \gamma^{\mu(\rho}\gamma^{\sigma)\nu}\, .\eqn{(4.22)}\fe

\bigskip
\noindent
5 SMALL GRAVITATIONAL PERTURBATIONS OF AN ELASTIC MEDIUM

\medskip

Having seen how to set up a system of exactly self - consistent but
non - linear equations governing a perfect elastic medium in the
framework of General Relativity, we are now ready to derive the
linearised wave equation governing small perturbations relative to
a known background, such as might be induced by weak incoming
gravitational radiation.

It is usually convenient to think of the perturbations as being
determined in terms of a vector field $\xi^\mu$ that specifies the
infinitesimal displacement of the worldlines relative to their
positions in the known background space. Of course such a displacement
is entirely gauge dependent and can always be reduced to zero by the
use of an appropriate Lagrangian (worldline dragging) mapping of the
perturbed space onto the background, butit is often convenient to fix 
the gauge by a more purely geometric requirement such as the
preservation of a harmonic coordinate system, which is the generalisation 
of the usual newtonian procedure of defining the displacements  relative 
to the fixed Euclidean structure of space. We shall use the symbol 
$\delta$ to denote the Eulerian variation of any quantity as specified
by any such geometric prescription for the mapping of the perturbed
spacetime onto the unperturbed background. The difference between the
Lagrangian variations denoted by $\Delta$ and the Eulerian variations
denoted by $\delta$ is given by Lie differentiation with respect to the
corresponding infinitesimal displacement field $\xi^\mu$, i.e.
\be \Delta=\delta +\xi\Libra \, .\eqno{(5.1)}\fe 
Thus in particular the Lagrangian variation of the metric tensor is given
by
\be \Delta_{\mu\nu}= h_{\mu\nu}+ 2\nabla_{(\mu}\xi_{\nu)}\, ,
\eqn{(5.2)}\fe
where we use the usual notation
\be h_{\mu\nu}=\delta g_{\mu\nu} \eqno{(5.3)}\fe
for the Eulerian variation of the metric arising from the gravitational
waves under consideration.

As far as quantities characterising the material medium are concerned,
it is easier to work with Lagrangian than Eulerian variations, since the 
former are related directly (via\ref{(3.13)}) to convected variations
and hence to the material variations that are governed directly by the
equations of state. In the case of orthogonal covariant tensors, 
including the special case of scalars, we see from \ref{(3.13)} that
the Lagrangian variation is given directly by the convected variation,
and hence from \ref{(4.12)} we find that the Lagrangian variation of
the density is given in terms of that of the metric by
\be \Delta\rho= -{_1\over^2}\rho y^{\rho\sigma}\Delta_{\rho\sigma}
\, ,\eqn{(5.4)}\fe
where for future convenience we introduce the abbreviation
\be y^{\rho\sigma}=\gamma^{\rho\sigma}+\rho^{-1}P^{\rho\sigma}\, .
\eqn{(5.5)}\fe
For a contravariant tensor however, \ref{(3.13)} introduces extra
terms, so that for the dependence of the Eulerian variation of the
pressure tensor in terms of the metric \ref{(4.14)} gives rise to the
formula
\be \Delta P^{\mu\nu}=-{_1\over^2}\big( E^{\mu\nu\rho\sigma}
+P^{\mu\nu}\gamma^{\rho\sigma} - 4 P^{\rho(\mu} u^{\nu)}u^\sigma\big)
\Delta_{\rho\sigma}\, .\eqn{(5.6)}\fe

Using these results together with the expression \ref{(3.3)} for the
lagrangian variation of the flow field $u^\mu$ itself, we are now in
a position to work out the perturbed equations of motion by taking the
variation of the exact equations of motion \ref{(4.7)}. It is evidently
most convenient to start from the Lagrangian variation, i.e.
\be \Delta\big(\rho\dot u^\mu+\gamma^\mu_\nu\nabla_{\!\rho}P^{\nu\rho}
\big) = 0\, ,\eqn{(5.7)}\fe
which works out explicitly as
\be \big(A^{\mu(\nu\ \sigma)}_{\ \ \ \, \rho}-\rho y^\mu_{\ \rho} u^\nu 
u^\sigma\big) \Delta\Gamma^\rho_{\ \nu\sigma}+\gamma^\mu_{\ \rho}
\epsilon_{\nu\sigma} \Delta_{\!\tau}E^{\rho\tau\nu\sigma}\hskip 3 cm\fe
\be =\Big(P^{\mu\nu}\dot u^\sigma-{_1\over^2}\dot u^\mu P^{\nu\sigma}
-2 A^{\mu(\nu\ \tau)}_{\ \ \,\rho}(\theta^\rho_{\ \tau}+
\omega^\rho_{\ \tau}) u^\sigma +\rho y^\mu_{\ \rho} \dot u^\rho u^\nu 
u^\sigma\Big)\Delta_{\nu\sigma} \eqn{(5.8)}\fe
using the abbreviation
\be \epsilon_{\mu\nu} = {_1\over^2}\delta \gamma_{\mu\nu}={_1\over^2}
\gamma^\rho_{\ \mu}\gamma^\sigma_{\ \nu}\Delta_{\rho\sigma} 
\eqn{(5.9)}\fe
for the relative strain tensor, and
\be A^{\mu\nu\ \sigma}_{\ \ \, \rho}=E^{\mu\nu\ \sigma}_{\ \ \,\rho}
-\gamma^\mu_{\ \rho} P^{\nu\sigma}   \eqn{(5.10)}\fe
for the modified elasticity tensor first introduced in a classical
context by Hadamard. When the Lagrangian metric perturbation 
$\Delta_{\mu\nu}$ is evaluated by use of \ref{(5.2)}, and the
corresponding perturbation of the affine connection components 
$\Gamma^\mu_{\ \nu\sigma}$ is evaluated using the corresponding 
substitution
$$\Delta \Gamma^\mu_{\ \nu\sigma} =\nabla_{\!(\sigma}
\Delta^\mu{_{\! \nu)}}-{_1\over ^2}\nabla^\mu\Delta_{\nu\sigma}\hskip 4 cm$$
\be \hskip 3 cm =\nabla_{\!(\nu}\nabla_{\!\sigma)}\xi^\mu
+\nabla_{(\nu} {h^\mu}{_{\!\sigma)}}-{_1 \over ^2}\nabla^\mu
h_{\nu\sigma} -\xi^\rho R^\mu_{\ (\nu\sigma)\rho}\eqn{(5.11)}\fe
where the Riemann tensor is defined by
\be \big(\nabla_{\!\mu}\nabla_{\!\nu}-\nabla_{\!\nu}\nabla_{\!\mu}
\big)\xi_\rho= R_{\mu\nu\rho\sigma}\xi^\sigma\, ,\eqn{(5.12)}\fe
then we see that the basic perturbation equation \ref{(5.8)} takes
the form of a hyperboliv wave equation for the displacement vector
$\xi^\mu$ when $h_{\mu\nu}$ is given. The characteristic cones and
the corresponding sound speeds can be worked out directly from
\ref{(4.7)} by considering discontinuities\cite{[10]} without any
need of the full set of perturbation equations given here.

In order to have a complete system of equations governing the
interaction of weak gravitational radiation with an elastic medium
we need an additional wave equation governing the gravitational
perturbation $h_{\mu\nu}$. This is obtainable by taking the
appropriate perturbations of the Einstein gravitational equations,
which are expressible (in units with {\Srm G}=1) by
\be \hat R^{\mu\nu}=8\pi T^{\mu\nu}  \eqn{(5.13)}\fe
where the Ricci tensor is defined by
\be R_{\mu\nu}=R_{\mu\rho\nu}^{\ \ \ \,\rho} \eqn{(5.14)}\fe
and where we use the notation $\hat{}$ for the partial trace 
subtraction operation defined by
\be \hat R^{\mu\nu}= R^{\mu\nu}- {_1\over ^2}g^{\mu\nu} R_\rho^{\ \rho}
\, .\eqn{(5.15)}\fe

In our work up to this stage we have been concentrating on material 
aspects, so that Lagrangian variations have given the simplest 
formulae. However the advantages of being able to use more general 
Eulerian variations become apparent now that we come to the properly
gravitational aspect, since it is well known that the perturbed
Einstein equations in the Eulerian form
\be \delta\big( \hat R^{\mu\nu}-8\pi T^{\mu\nu}\big)=0
\eqn{(5.16)}\fe
can be greatly simplified by the imposition of the harmonic
gauge condition
\be \nabla_{\!\mu}\hat h^{\mu\nu}=0\, .\eqn{(5.17)}\fe
Under these conditions most of the terms drop out and one is left with
an equation of the form
\be \Square \hat h^{\mu\nu}= - 16\pi\delta T^{\mu\nu}\eqn{(5.18})\fe
where the relevant wave operator is defined by
\be \Square \hat h^{\mu\nu}=\nabla_\rho\nabla^\rho \hat h^{\mu\nu}
-\hat R^{\mu\nu} \hat h_\rho^{\ \rho}+ 2C^{\mu\ \nu}_{\ \rho\ \sigma}
\hat h^{\rho\sigma}\fe
\be \hskip 3 cm -{_2\over^3}R_\rho^{\ \rho}\big(\hat h^{\mu\nu}
-{_1\over^4} g^{\mu\nu}\hat h_\rho^{\ \rho}\big) \eqn{(5.19)}\fe
using the standard notation
\be C^{\mu\nu}_{\ \ \rho\sigma}=R^{\mu\nu}_{\ \ \rho\sigma}- 2
g^{[\mu}_{\,[\rho}\big( R^{\nu]}_{\,\sigma]}-{_1\over^6}
g^{\nu]}_{\,\sigma]} R^\tau_{\ \tau}\big) \eqn{(5.20)}\fe
(with square brackets denoting antisymmetrisation) for Weyl's
trace free conformal tensor. The Eulerian variation of the 
energy-momentum tensor is obtainable using \ref{(5.1)} in terms
of a Lie derivative and a more easily evaluable Lagrangian
variation. Thus starting from \ref{(4.1)} and using \ref({3.1}),
\ref{(5.4)}, and \ref{(5.6)} we obtain finally
$$ \delta T^{\mu\nu}= -{_1\over^2}({\cal E }^{\mu\nu\rho\sigma}
 +T^{\mu\nu} g^{\rho\sigma}) \Delta_{\rho\sigma} $$
\be \hskip 1cm +2 T^{\rho(\mu}\nabla_{\!\rho}\xi^{\nu)}
-\xi^\nu\nabla_{\!\rho} T^{\mu\nu} \eqn{(5.21)}\fe
where, following Friedman and Schutz\cite{[14]} we have constructed
a generalised (non-orthogonal) elasticity tensor with the same
symmetry properties as those of the ordinary elasticity tensor
\ref{(4.16)} according to the prescription
\be {\cal E}^{\mu\nu\rho\sigma}= E^{\mu\nu\rho\sigma}+
6 u^{(\mu} u^\nu P^{\rho\sigma)}- 8 u^{(\mu} P^{\nu)(\rho} 
u^{\sigma)}-\rho u^\mu u^\nu u^\rho u^\sigma\, .\eqn{(5.22)}\fe

The coupled system of equations \ref{(5.8)} and \ref{(5.9)}
simplifies considerably when not only the perturbations but also
the background gravitational field is weak, as is the case in
terrestrial (as opposed to neutron star) applications. In such
cases we may suppose that there is a small dimensionless
parameter, $\epsilon$, loosely interpretable as an upper bound
not only on the order of magnitude of the gravitatinal wave 
perturbations $h_{\mu\nu}$, $\epsilon_{\mu\nu}$ etc., but also
of the deviations of the metric from the flat Minkowski form.
It then follows from the Einstein equation that the density $\rho$
must also be of linear order in $\epsilon$ as the latter tends
to zero, while (by the virial theorem) the pressure in a self
gravitating system is of even higher order, tending to zero
even when divided by $\epsilon$, i.e. in standard notation
\be P^{\mu\nu}= o\{\epsilon\} \, .\eqn{(5.23)}\fe
(In our original version \cite{[11]} a printer's error substituted
0 in place of o throughout.) Since the unperturbed energy-momentum
tensor will be at most of linear order in $\epsilon$ its
perturbation will be of higher order, i.e.
\be \delta T^{\mu\nu}= o\{\epsilon\} \eqn{(5.24)}\fe
so that the gravitational wave equation \ref{(5.18)} will to lowest
order be of simple Dalembertian form , i.e.
\be \nabla_{\!\rho}\nabla^{\!\rho} h^{\mu\nu}= o\{\epsilon\}
\, .\eqn{(5.25)}\fe
It follows that in addition to the harmonic gauge condition we can to 
this order make the further simplifications
\be h_{\mu\nu} u^\nu= o\{\varepsilon\}\, ,\hskip 1 cm
h^\nu_{\,\nu}= o\{\epsilon\} \eqn{(5.26)}\fe
and to the same order the wave equation for the displacement
reduces to the form
\be u^\nu u^\sigma \nabla_{\!\nu}\nabla_{\!\sigma}\xi^\mu -
\rho^{-1} \nabla_{\!\nu}\big(E^{\mu\nu\rho\sigma}(\nabla_{\!\rho}
\xi_\sigma+{_1\over^2}h_{\rho\sigma})\big)=o\{\epsilon\}
\eqn{(5.27)}\fe
in agreement with calculations of Dyson \cite{[2]} and Papapetrou
\cite{[3]} .

As a simple practical application Dyson considered the case of an elastic 
medium that is $\rm\underline{isotropic}$, as it will be in the case of a
typical metal when considered on scales large compared with the
microscopic crystalline domains. In such a case (in consequence of
\ref{(5.23)}) the elasticity tensor will be of the form
\be E^{\mu\nu\rho\sigma}=\beta \gamma^{\mu\nu} \gamma^{\rho\sigma}
+ 2\mu \big( \gamma^{\mu(\rho}\gamma^{\sigma)\nu} -{_1\over^3}
\gamma^{\mu\nu}\gamma^{\rho\sigma}\big) \eqn{(5.28)}\fe
where $\beta$ is the bulk modulus and $\mu$ is the rigidity
modulus. Now under these conditions it follows from
\ref{(5.26)} that the only gravitational term in \ref{(5.28)}
reduces to the form
\be \nabla_{\!\nu}\big(E^{\mu\nu\rho\sigma} h_{\rho\sigma}\big)
=2 h^{\mu\nu} \nabla_{\!\nu}\mu \, ,\eqn{(5.29)}\fe
which shows, as pointed out by Dyson, that the gravitational
waves couple with the elastic displacement only via non-uniformities
of the rigidity. (In the case of a traditional Weber bar detector
the relevant non-uniformity is provided by the discontinuity at
the surface of the cylinder.)

\bigskip
REFERENCES
\medskip
\parindent = 0 cm

[1] Weber, J., {\it Phys. Rev.} {\bf 117} (1960) 306.

[2] Dyson, F.J., {\it Astroph. J.} {\bf 156} (1969) 529.

[3] Papapetrou, A. {\it Ann. Inst. H. Poincar\'e} {\bf A16}
(1972) 63.

[4] Carter B., and Quintana, H., {\it Astroph. J.} {\bf 202}
(1975) 54.

[5] Maugin, G.A., {\it J. Math. Phys.} {\bf 19} (1978) 1198.

[6] Souriau, J.M.,  {\it G\'eom\'etrie et Relativit\'e}
(Herman, Paris, 1965).

[7] Oldroyd, J.G., {\it Proc. Roy. Soc.} {\bf A270}
(1970) 103.

[8] Carter, B., and Quintana, H., {\it Proc. Roy. Soc.} {\bf A331},
(1972) 57.

[9] Carter, B., {\it Commun. Math. Phys.} {\bf 30} (1973) 261.

[10] Carter, B., {\it Phys. Rev.} {\bf D7} (1973) 1590.

[11] Carter, B., and Quintana, H., {\it Phys. Rev.} {\bf D16}
(1977) 2928.

[12] Carter, B., {\it Proc. Roy. Soc.} {\bf A372} (1980) 169.

[13] Ehlers, J., in Israel, W., (ed) {\it Relativity, Astrophysics
and Cosmology} (Reidel, Dordrecht, 1970) 89.

[14] Penrose, R., in DeWitt, C., and Wheeler, J.A. (eds) {\it Battelle
Rencontre} (Gordon and Breach, New York, 1968)

[15] Friedman, J.L., and Schutz, B.L., {\it Astrophys. J.} {\bf 200}
(1975) 204.

\end